\documentclass[prb,twocolumn,superscriptaddress]{revtex4}
\usepackage{amsmath,amssymb,graphicx,color,bm,epstopdf}

\begin{document}

\title{Magnetic properties of graphene quantum dots}
\author{T. Espinosa-Ortega}
\affiliation{Instituto de Energ\'{\i}as Renovables, Universidad Nacional Aut\'onoma de
M\'exico, Temixco, Morelos, 62580, M\'exico} \affiliation{Division of Physics and Applied
Physics, Nanyang Technological University 637371, Singapore}

\author{I. A. Luk'yanchuk}
\affiliation{LPMC,  University of Picardie
 Amiens, France and Landau Institute for Theoretical Physics, Russia}

\author{Y. G. Rubo}
\affiliation{Instituto de Energ\'{\i}as Renovables, Universidad Nacional Aut\'onoma de
M\'exico, Temixco, Morelos, 62580, M\'exico}

\begin{abstract}
Using the tight-binding approximation we calculated the magnetic
susceptibility of graphene quantum dots (GQDs) of different geometrical
shapes and characteristic sizes of 2-10 nm, when the magnetic properties are
governed by the electron edge states. Two types of edge states can be
discerned: the zero-energy states (ZES) located exactly at the zero-energy
Dirac point, and the dispersed edge states (DES) with the energy close, but
not exactly equal to zero. DES are responsible for the temperature
independent diamagnetic response, while ZES provide the temperature
dependent spin Curie paramagnetism. The hexagonal, circular and randomly
shaped GQD contain mainly DES and, as a result, they are diamagnetic. The
edge states of the triangular GQD are of ZES type. These dots reveal the
crossover between spin paramagnetism, dominating for small dots and at low
temperatures, and orbital diamagnetism, dominating for large dots and at
high temperatures.
\end{abstract}

\maketitle

\affiliation{Instituto de Energ\'{\i}as Renovables, Universidad Nacional Aut\'onoma de
M\'exico, Temixco, Morelos, 62580, M\'exico}
\affiliation{Division of Physics and Applied
Physics, Nanyang Technological University 637371, Singapore}

\affiliation{LPMC, University of Picardie Amiens, France and Landau Institute for
Theoretical Physics, Russia}

\affiliation{Instituto de Energ\'{\i}as Renovables, Universidad Nacional Aut\'onoma de
M\'exico, Temixco, Morelos, 62580, M\'exico}


\section{Introduction}

\label{sec:intro}

In past years the special attention has been paid to fabrication of graphene
quantum dots (GQDs) \cite{Bunch,Ozylmaz} susceptible to be used for magnetic
field-controlled spin-electronic logic gates \cite{WeiWang}. However the
origin of magnetism in such structures still remains unclear in spite of
number of recent studies of orbital \cite%
{Zhang,Schnez,Potasz,Zarenia,Grujic,LukBrat} and spin \cite%
{Ezawa,Rossier,Wang} properties. This concerns, in particulary, the
dependence of magnetic response on the size and on the shape of GQD.

Landau diamagnetism in perfect infinite graphene sheet was first studied by
McClure \cite{McClure1,McClure2,McClure3} and more recently in Refs.~[%
\onlinecite{Fukuyama,Nakamura,Ghosal,Slizovskiy,Sharapov}], were the
singular behavior of susceptibility was found when the Fermi energy
approaches the Dirac point at zero temperature. This peculiar behavior
around zero energy also takes place in the cases when disorder-provided band
is present for infinite graphene and ribbons.\cite{Waka,Koshino1,Liu,Koshino2,Ominato} On the other hand,
the presence of the edge states with energy around zero is a signature of
the graphene nanoflakes with various terminations and most notably with the
zig-zag edges.\cite{Nakada,Fujita,Wurm}

The number and the properties of edge states are sensitive to the geometry
of the GQD.\cite{Heiskanen,Espinosa,Rozhkov} Since diamagnetism of graphene
occurs due to the electronic states with the energy near the Dirac point, it
is natural to assume that the edge states should make a dominant
contribution to magnetism of graphene nanoflakes and the geometry of GQD
will play an important role in the diamagnetic response of the nanostructure.

In this paper we study the hexagonal, circular, triangular and random GQDs
and identify two types of the edge states. First, there are the dispersed
edge states (DES) whose energies are distributed in the range of $2\Delta$
around the Dirac point, with the value of $\Delta$ being inversely
proportional to the size of the GQD. Secondly, there could be highly
degenerate exactly-zero-energy states (ZES). The DES are appropriate to the
hexagonal, circular and random GQDs. Their energies are sensitive to the
applied field that induces the edge currents. These states provides the
orbital diamagnetic response of the nanoflakes. The number of ZES, that are
mostly present in triangular GQDs, can be found exactly from the graph
theory \cite{Fajtlowicz}. Their origin is purely geometric and their
location does not change as function of the applied magnetic field.
Therefore ZES do not contribute to the diamagnetism of GQDs, but they can be
occupied by the electrons with unpaired spins and provide the paramagnetism
of the system \cite{Rossier}. Studying the edge-state-provided
orbital-diamagnetic and spin-paramagnetic response of GQD we predict the
possibility of crossover between paramagnetic and diamagnetic response of
GQD as a function of their shape, size and temperature.

After this work was completed, we became aware of the preprint \cite%
{Ominato2} where the similar research was done. The main difference between
our results is that, we consider the dots of smaller size and at low
temperatures $k_{B}T\ll \Delta $, where the edge-states-provided diamagnetic
peak is broaden by the size effects and is temperature independent.   Ref.~[%
\onlinecite{Ominato2}] mainly addresses the temperature effects relevant for
the GQDs with bigger sizes and small values of $\Delta $ when diamagnetism
is mainly of the bulk origin.

\section{The model of graphene quantum dots}

\label{sec:Model}

We use the simplest nearest-neighbor tight-binding approximation where the
properties of conducting $\pi $-electrons of graphene are described by the
Hamiltonian
\begin{equation}
H=\sum_{i}\varepsilon _{i}c_{i}^{\dag }c_{i}+\sum_{\left\langle
ij\right\rangle }t_{ij}c_{i}^{\dag }c_{j},  \label{Htb}
\end{equation}%
where $c_{i}^{\dag }$, $c_{i}$ are the creation and annihilation electron
operators and $\varepsilon _{i}$ is the on-site energy. In what follows we
do not consider any on-site disorder and set $\varepsilon _{i}=0$. The
hopping matrix elements $t_{ij}$ between nearest-neighbor carbon atoms
account for the magnetic field via the Peierls substitution,
\begin{equation}
t_{ij}=\gamma _{0}\exp \left\{ \frac{e}{\hbar c}\int_{\mathbf{r}_{i}}^{%
\mathbf{r}_{j}}\mathbf{A}\cdot d\mathbf{l}\right\} ,  \label{Hopping}
\end{equation}%
where $\mathbf{A}=(0,Bx,0)$ is the vector potential of magnetic field and
the zero-field hopping was taken as $\gamma _{0}=3.0\,\mathrm{eV}$.

The graphene flakes were selected of hexagonal, circular, triangular and
random shapes with mostly zig-zag edges. The contour of the random shape
nanostructures has been defined in polar coordinates by:
\begin{equation}
r(\theta )=r_{0}+\sum_{k=1}^{k_{\mathrm{max}}}[A_{k}\cos (k\theta
)+B_{k}\sin (k\theta )]  \label{rrand}
\end{equation}%
where $r_{0}$ is the constant average radius that defines the typical size
of the flakes, $A_{k}$ and $B_{k}$ are the random numbers with amplitude not
exceeding $r_{0}/3$. In order to have realistic variation of the flake edge
on the scale of the lattice constant, the maximum number of harmonics $k_{%
\mathrm{max}}$ has been chosen to be of the order of $r_{0}/a$, where $%
a=2.461\,\mathrm{\mathring{A}}$ is the lattice constant. Typically, this
number was about 25.

Direct numerical diagonalization of Hamiltonian (\ref{Htb}) gives the
field-dependent energy levels $E_{n}(B)$ and corresponding on-site
amplitudes $\varphi _{n,i}$ of the wave function. The orbital energy of the $%
\pi $-electrons at zero temperature as a function of the chemical potential $%
\mu $ and magnetic field $B$ is given by
\begin{equation}
U(B,\mu )=2\sum_{n}^{E_{n}<\mu }E_{n}(B),  \label{TotalE}
\end{equation}%
where the factor of 2 is the spin-degeneracy of the levels. The
low-temperature diamagnetic susceptibility per unit area has been calculated
as
\begin{equation}
\chi (\mu )=-\frac{1}{\sigma }\left[ \frac{\partial ^{2}U(B,\epsilon )}{%
\partial B^{2}}\right] _{B=0},  \label{Sus}
\end{equation}
where $\sigma =\sqrt{3}a^{2}N/4$ is the area of graphene flake containing $N$
carbon atoms.

\section{Bulk and edge states}

\label{sec:EdgeStates}

\begin{figure}[h]
\includegraphics[width=7cm,height=13.5cm]{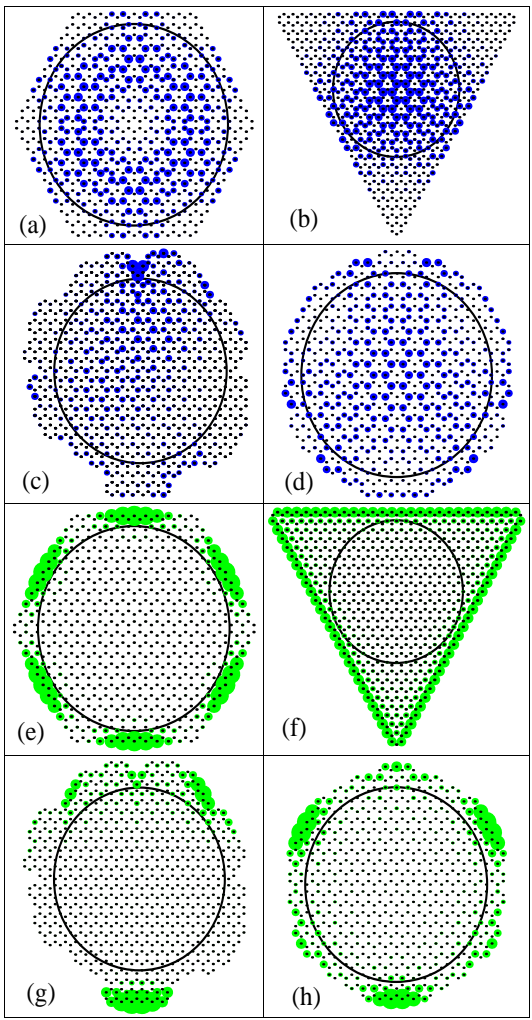}
\caption{Amplitude of the wave function for bulk (a-d) and edge
(e-h) states in GQD of different shapes. We also show the maximal
inscribed circles that have been used to discern between bulk and
edge states.}
\label{Fig1}
\end{figure}

In what follows it will be convenient to distinguish the bulk and the edge
electronic states using the following geometrical criterium. For a given
state with the energy $E_{n}$ we ascribe the intensity $I_{n}^{(b)}=%
\sum_{r_{i}<R}|\varphi _{n,i}|^{2}$ of the electronic states located within
the circle of radius $R$ to the \textit{bulk} part of the total wave
function intensity, whereas the outer part $I_{n}^{(e)}=\sum_{r_{i}>R}|%
\varphi _{n,i}|^{2}=1-I_{n}^{(b)}$ will be due to the \textit{edge}
contribution. The radius $R$ has been chosen to be about of one lattice
constant smaller than the radius of the maximum circle that can be inscribed
in a given GQD. Then, the state will be referred to as the edge state if $%
I_{n}^{(e)}>I_{n}^{(b)}$. Otherwise, we refer to it as to the bulk state.
Wave functions of the typical edge and bulk states are illustrated in Fig.~%
\ref{Fig1}.

\begin{figure}[h]
\includegraphics[width=8cm,height=8cm]{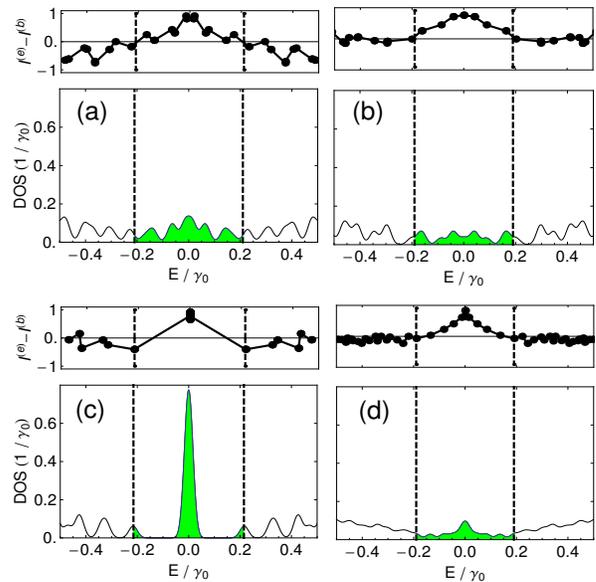}
\caption{ Electronic density of states and the difference of the edge and
the bulk contributions intensities $I_n^{(e)}-I_n^{(b)}$ (see text) as a
function of energy for the hexagonal (a), circular (b), triangular (c) and
random (d) GQDs. The shaded region indicates the pseudogap where the edge
states are located. The levels have been artificially shown as Gaussian
peaks with the dispersion $0.05\mathrm{eV}$.}
\label{Fig2}
\end{figure}

The bulk and the edge states distinguished by the above criterium are also
separated in energy. Namely, the edge states normally possess the energy $%
|E|<\Delta $, while the energy of bulk states $|E|>\Delta $. We will refer
the energy interval of $2\Delta $ around the Dirac point as \emph{pseudogap}%
. It turns out that the value of pseudogap is approximately equal for all
GQDs, characterized by the same inner radius $R$ (see Fig.~\ref{Fig2}). The
pseudogap scales as $\Delta \propto \gamma_{0}/\sqrt{N}\propto \gamma _{0}a/R
$.

Two types of the edge states can be discerned: (i) the zero-energy states
(ZES) that are degenerated and located exactly at $E=0$, i.e. in the middle
of the pseudogap, and (ii) the dispersed edge states (DES) that have the
non-zero energies, are symmetrically distributed with respect to $E=0$, and
fill the pseudogap.

As it was shown by the graph-theory \cite{Fajtlowicz} the total number of
ZES is related to the imbalance between the $A$ and $B$-type atoms in the
graphene flake:
\begin{equation}
\eta _{0}\geq |N_{A}-N_{B}|,  \label{disbal}
\end{equation}%
where the equality is taking place for the geometry of equilateral polygons.
For hexagons $\eta _{0}=0$, there are no ZES and all edge states are of the
DES type. Contrary, for the equilateral triangles all the edge states are of
ZES type and their degeneracy number is given by \cite{Potasz,Zarenia}
\begin{equation}
\eta _{0}^{\Delta }=\sqrt{N+3}-3.  \label{DegTri}
\end{equation}%
Usually there are only few ZES for circular and randomly shaped GQDs.

The number of ZES does not depend on magnetic field and therefore these
levels do not contribute to the orbital part of susceptibility (\ref{Sus}).
In contrast, they are responsible for the spin-provided super-paramagnetic
response of ensemble of clusters in case of the half-filled $\pi $-band when
the Fermi energy is pinned at $\mu =0$. Indeed, according the Hund theorem,
the number of single occupied states of degenerate level should be maximal,
providing the total uncompensated spin defined by the Lieb's rule \cite%
{Rossier,Wang} $S=\frac{1}{2}\eta _{0}$ that brings the substantial
contribution to the temperature-dependent spin-Curie paramagnetism.

On the other hand the location of DES in hexagonal, circular and random GQD\
depends on the applied field and therefore these levels are responsible for
diamagnetism of graphene clusters, as it is calculated below.

\section{Hexagons, circles and random quantum dots}

\label{sec:DESDiamag}

The susceptibility was calculated for zig-zag edge hexagons and circles of
about of ten different sizes, having an inner radius in the range of $2-7\,%
\mathrm{nm}$. For random quantum dots, the averaging has been performed over
three different ensembles, characterized by mean inner radii $R_{1}=2.6\,%
\mathrm{nm}$, $R_{2}=3.65\,\mathrm{nm}$ and $R_{3}=4.69\,\mathrm{nm}$. The
magnetic field varied between 0 and $5\,\mathrm{T}$, range in which the
susceptibility remained approximately constant. All the plots are presented
for $B=5\,\mathrm{T}$. Fig.~\ref{Fig2} shows the density of states as a
function of Fermi energy for hexagonal and random-shape GQD. The shaded area
indicates the region where the edge states are located.

\begin{figure}[tbp]
\includegraphics[width=7.5cm,height=15cm]{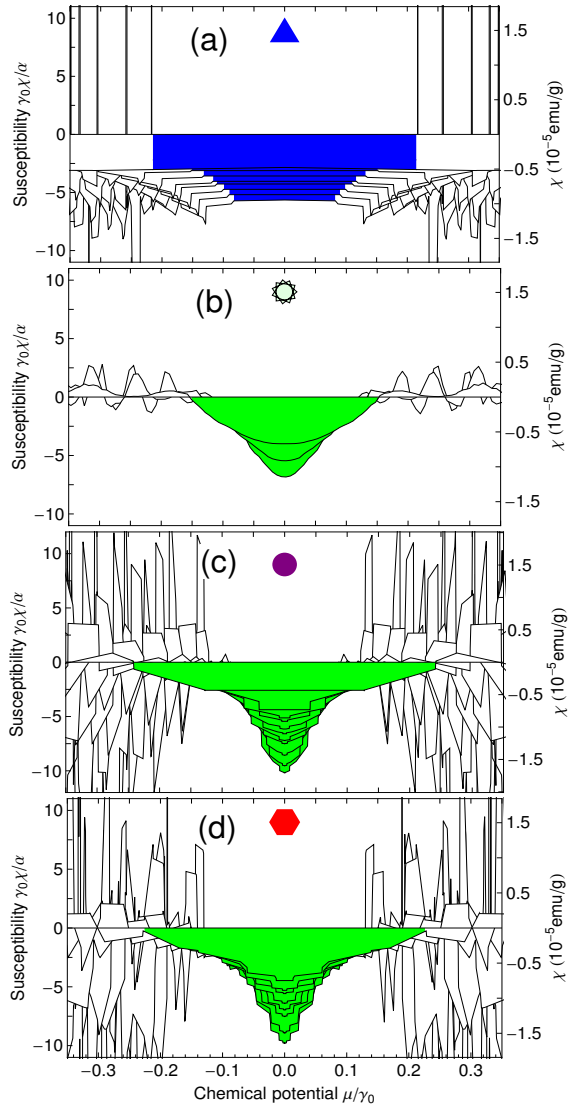}
\caption{Susceptibility for triangular (a), random (b), circular (c) and
hexagonal (d) GQDs of sizes $2-7\,\mathrm{nm}$. The shaded region indicates
the (pseudo)gap. }
\label{Fig3}
\end{figure}

Fig.~\ref{Fig3} shows the magnetic susceptibility per unit of area as
function of Fermi energy for GQD of different sizes. As was qualitatively
explained above, the diamagnetic peak of width $2\Delta$ appears when the
chemical potential crosses the pseudogap. This peak becomes wider with
decreasing of GQD size. Beyond this zone, the orbital susceptibility is a
highly fluctuating function of the Fermi energy that oscillates between
paramagnetic and diamagnetic sign. These oscillations have been recently
interpreted for graphene ribbons, as a result of the sub-band structure \cite%
{Ominato}.

As it was mentioned above, the number of ZES for these geometries is
vanishingly small that makes the diamagnetic contribution dominant for all
sizes.

\section{Triangular quantum dots}

\label{sec:Tri}

\begin{figure}[t]
\includegraphics[width=7.5cm,height=4cm]{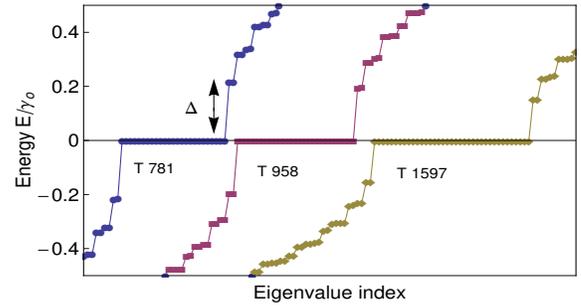}
\caption{The band gap for triangular GQD with different number of atoms $N$.}
\label{Fig4}
\end{figure}

All the edge states in triangular GQD with zig-zag edges are of ZES-type
that do not change their energy as function of the field and do not
contribute to the diamagnetic susceptibility. The ZES for GQD of different
sizes are located in the middle of the gap $2\Delta $ as is shown in Fig.~%
\ref{Fig4}. We note that $2\Delta $ indicates the real gap in the spectrum
for the triangular GQD, and the value of the this gap is inversely
proportional to the number of ZES $\eta_{0}^{\Delta }$:
\begin{equation}
\Delta =\frac{\zeta \gamma _{0}}{\eta _{0}^{\Delta }} =\frac{\zeta \gamma
_{0}}{\sqrt{(N+3)}-3},  \label{tgap}
\end{equation}%
where the numerical constant is $\zeta \simeq 5.56$.

The magnetic susceptibility of triangular GQD $\chi _{orb}^{\Delta }$ is
shown in Fig.~\ref{Fig3} for nine inner sizes of $2.5-7\,\mathrm{nm}$. It is
provided by the out-of-gap delocalized electronic states and does not depend
on $\mu $ within the gap $2\Delta $ because of absence of DES.\ These
results match the analytical calculations of Ref.~[%
\onlinecite{Sharapov,Koshino2}] for an infinite graphene sheet with the band
gap $2\Delta $, according to which the diamagnetic susceptibility per unit
area is
\begin{equation}
\chi _{\mathrm{orb}}^{\Delta }(\mu )=-\alpha \frac{\theta (\Delta -|\mu |)}{%
2\Delta },\qquad \alpha =\frac{e^{2}\gamma _{0}^{2}a^{2}}{2\pi \hbar
^{2}c^{2}},  \label{xdelta}
\end{equation}%
where $\theta (x)$ is the step function.

Although ZES give no contribution to the orbital susceptibility, they can be
responsible for the huge paramagnetism provided by $\eta _{0}$ uncompensated
electron spins located on the degenerate ZES levels. This happens in the
case of small positive chemical potentials if the energy of
electron-electron repulsion in each zero-energy state $U_{e-e}>\mu$, so that
these levels remain half-filled. The corresponding Curie-type
temperature-dependent paramagnetic susceptibility for non-interacting
electrons is evaluated per unit area as
\begin{equation}
\chi_{\mathrm{spin}}^{\Delta }=\frac{\eta _{0}^{\Delta }}{\sigma } \frac{%
(g\mu _{B})^{2}}{3}\frac{s(s+1)}{k_\mathrm{B}T} =\frac{\eta _{0}^{\Delta}\mu
_{B}^{2}}{\sigma k_\mathrm{B}T},  \label{Xspin}
\end{equation}%
where $\mu _{B}$ is the Bohr magneton and $g\simeq 2$ is the g-factor of the
electrons with spin $s=1/2$.

In the opposite case of the strong Coulomb electron correlations, according
to the Lieb theorem\cite{Lieb}, all $\eta _{0}^{\Delta }\gg 1$ ZES electrons form
the total spin of the cluster $S=\eta_{0}/2$. The super-paramagnetic
susceptibility of ensemble of triangular GQD becomes even stronger,
\begin{equation}
\chi _{\mathrm{spin}}^{\prime \Delta } =\frac{1}{\sigma }\frac{(g\mu
_{B})^{2}}{3}\frac{S(S+1)}{k_\mathrm{B}T} \simeq \frac{\left( \eta
_{0}^{\Delta }\mu_{B}\right) ^{2}}{3\sigma k_\mathrm{B}T}.  \label{XS}
\end{equation}
The actual value of paramagnetic susceptibility should be somewhere in
between of $\chi_{\mathrm{spin}}^{\Delta }$ and $\chi _{\mathrm{spin}%
}^{\prime \Delta }$.

Using Eqs.~(\ref{Xt}) and (\ref{XS}) we compare the spin-paramagnetic and
orbital-diamagnetic contributions, presenting their ratio for the case of
strongly correlated electrons as
\begin{equation}
\left\vert \frac{\chi_\mathrm{spin}^{\prime \Delta }}{\chi_\mathrm{orb}%
^{\Delta }}\right\vert \simeq \frac{7.1eV}{\sqrt{N}k_\mathrm{B}\,T}.
\label{ratio}
\end{equation}%
It follows that varying the size and temperature of the triangular quantum
dots, we can expect the paramagnetic-diamagnetic crossover. In particular,
for a temperature of $T=77\,\mathrm{K}$, we predict that triangular quantum
dots will be paramagnetic for the inner radius below $R\simeq 97\,\mathrm{nm}
$.

\section{Size Dependence}

\label{sec:Size}

We calculated the gap-zone-integral $\int_{-\Delta }^{\Delta }\chi (\mu
)d\mu $ of the shown in Fig. \ref{Fig3} diamagnetic susceptibilities for GQD
of different shape and found that it preserves its intensity while the width
$2\Delta $ vanishes as the size of GQD increases. In the limit of infinite
cluster with $\Delta \rightarrow 0$ this gives the McClure $\delta $-peak of
the graphene orbital susceptibility \cite{McClure1}:
\begin{equation}
\chi (\mu )=-\alpha \delta (\mu ).  \label{deltapeak}
\end{equation}

\begin{figure}[tbp]
\includegraphics[width=7.8cm,height=5cm]{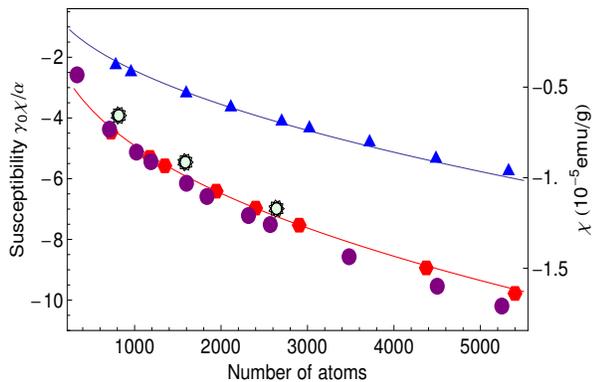}
\caption{Pseudogap $\Delta $ dependence on the number of atoms in GQD of
different geometry. The inset shows the energy dependence of susceptibility
for a circular GQD with $N=709$}
\label{FigDelta}
\end{figure}

The shown in Fig.\ref{FigDelta} dependence of $\Delta $ on the number of
atoms, is very similar for all shapes and follow the Eq.(\ref{tgap})

\begin{figure}[h]
\includegraphics[width=8.5cm,height=5cm]{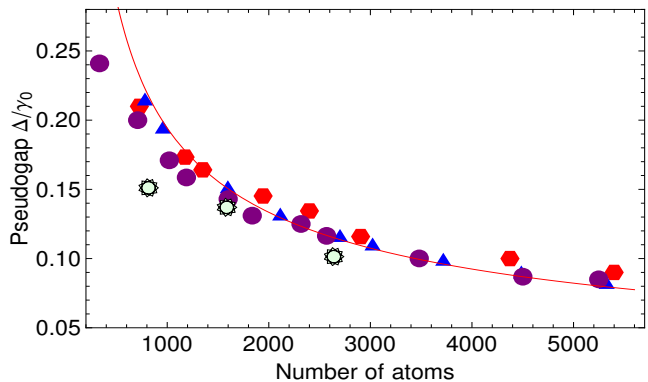}
\caption{Dirac-point orbital susceptibility at $\protect\mu =0$. The solid
lines correspond to the fit according Eqs.~(\protect\ref{Xt}) and (\protect
\ref{Xh}). }
\label{FigSize}
\end{figure}

The size-dependence of the orbital susceptibility for hexagonal, circular
and random GQDs at $\mu =0$ is shown in Fig.\ref{FigSize}. It satisfies an
empirical relation
\begin{equation}
\chi _{orb}=\beta N^{\lambda }  \label{Xh}
\end{equation}%
with $\lambda =0.4$ and $\beta =0.31\alpha /\gamma _{0}$. Although the
orbital diamagnetism is originated from the edge currents of the low-energy
DES, their perimeter contribution $\sim N^{1/2}$ can be reduced by the
armchair or/and zigzag type boundary irregularities as well as by the wave
function vanishing at the corners of GQD that explains the reduction of the
exponent index $\lambda $ slightly below $1/2$.

For triangular GQD the size-dependence of the orbital susceptibility can be
obtained by substituting the value of $\Delta $ from Eq.(\ref{tgap}) into
Eq.(\ref{xdelta}) at $\mu =0$:
\begin{equation}
\chi _{\mathrm{orb}}^{\Delta }=-\frac{\alpha }{\gamma _{0}}\frac{\sqrt{N+3}-3%
}{2\zeta }.  \label{Xt}
\end{equation}%
This estimation agrees with our numerical calculations as shown in Fig.\ref%
{FigSize}.

So far, there are no experimental measurements of magnetism of GQD but the
susceptibility we obtained is in the range of the measured values for carbon
nanotubes and buckyballs \cite{Bandow}. For convenience we also presented
the susceptibilities in Figs.~\ref{Fig3} and \ref{FigSize} in the
experimental units of emu/g.

\section{Conclusions}

Magnetism of GQD is provided by the edge states whose energy is located
within the finite-size quantization pseudogap. The structure of the edge
state spectrum and magnetic response of GQDs being strongly dependent on the
geometric shape of the cluster.

For hexagonal, circular and random GQD the edge states are dispersed within
the pseudogap. Their position depends on the applied field, providing the
substantial diamagnetic response of GQD. The diamagnetic susceptibility as
function of the chemical potential presents a peak of constant intensity,
centered around $\mu =0$. The maximum of the peak increases with GQD size
whereas its width decreases, approaching the $\delta $-function of McClure (%
\ref{deltapeak}) for infinite sheet of graphene.

For triangular GQD the edge states are located exactly at the middle of the
gap with the high degeneracy factor $\eta _{0}^{\Delta }$ given by Eq.(\ref%
{tgap}) that increases with size of the cluster. The zero-energy position of
these levels do not change with the field and the diamagnetic response of
triangular GQD $\chi _{\mathrm{orb}}^{\Delta }$ is expected to be small. In
a contrast, the uncompensated spins of electrons localized at ZES can
provide the huge paramagnetic temperature-dependent contribution $\chi _{%
\mathrm{spin}}^{\Delta }$ of the Curie type. By comparison of
susceptibilities $\chi _{\mathrm{orb}}^{\Delta }$ and $\chi_\mathrm{spin}%
^{\Delta } $, Eq.(\ref{ratio}) we expect to have the crossover from
paramagnetic to diamagnetic response in ensemble of triangular clusters with
increasing the temperature and/or the GQD size.

The strong dependence of magnetic properties of GQDs on their geometry, size
and temperature provides the natural way to separate the graphene clusters
according to their shape and size by application of the appropriately
designed non-uniform magnetic field and temperature cycle that can trap the
different GQD in different points of space. It would be interesting also to
study the specially cut nano-clusters of highly ordered pyrolytic graphite
that can contain the separate graphene sheets with Dirac-like spectrum \cite%
{LukDirac} and therefore can have the similar magnetic behavior.

\section*{Acknowledgments}

This work was supported in part by DGAPA-UNAM under the project No. IN112310
and by the EU FP7 IRSES projects POLAPHEN and ROBOCON. T. Espinosa-Ortega
thanks University of Picardie for hospitality where the part of this work
was done.


\end{document}